\documentclass[prl,aps,floatfix,twocolumn,superscriptaddress]{revtex4}

\usepackage{graphicx}
\graphicspath{{figures/}}

\usepackage{amsmath}
\usepackage{units,xspace}
\newcommand{\micron}{\ensuremath{\unit{\mu m}}\xspace}

\newcommand{\abs}[1]{\left\vert #1 \right\vert}
\newcommand{\avg}[1]{\left< #1 \right>}
\renewcommand{\vec}[1]{\ensuremath{\boldsymbol{#1}}}
\newcommand{\uvec}[1]{\ensuremath{\hat{\boldsymbol{#1}}}}

\begin{document}

\title{Brownian vortexes}

\author{Bo Sun}
\affiliation{Department of Physics and Center for Soft Matter
  Research, New York University, New York, NY 10003}

\author{Jiayi Lin}
\affiliation{NEST+m, 111 Columbia Street, New York, NY 10002}

\author{Ellis Darby}
\affiliation{NEST+m, 111 Columbia Street, New York, NY 10002}

\author{Alexander Y. Grosberg}

\author{David G. Grier}

\affiliation{Department of Physics and Center for Soft Matter
  Research, New York University, New York, NY 10003}

\begin{abstract}
A particle diffusing around a point of
stable mechanical equilibrium in a static but non-conservative force field
enters into a steady state characterized by circulation in the probability
flux.
Circulation in such a Brownian vortex is not simply
a deterministic response to the solenoidal component of the force, 
but rather reflects an interplay between force-driven probability 
currents and diffusion.
As an example of this previously unrecognized
class of stochastic heat engines,
we consider a colloidal sphere stably trapped in a conventional
optical tweezer.
Rather than coming into thermodynamic equilibrium with the
surrounding heat bath, the particle's Brownian fluctuations
are biased into a toroidal roll.
We demonstrate both theoretically and experimentally
that the circulation in this practical
realization of the Brownian
vortex can undergo flux reversal.
\end{abstract}


\maketitle

Stochastic heat engines such as thermal ratchets and Brownian
motors use non-conservative forces to eke a 
flux of energy or probability out of otherwise random thermal
fluctuations 
\cite{thermalratchets}.
Unlike deterministic machines, these systems would not operate at
all without the motivating influence of thermal noise.
Although a great many mechanisms have been proposed for implementing
stochastic heat engines, virtually all rely on time dependence
in the force field to rectify fluctuations.
Here, we demonstrate that time-independent force fields also
can create stochastic heat engines,
provided the force has a non-vanishing solenoidal component.
The resulting steady-state
probability currents
feature circulating rolls, which inspire us
to name such systems Brownian vortexes.
We demonstrate both theoretically and experimentally
that Brownian vortexes arise naturally in the motions
of colloidal
spheres localized in optical traps, and that the 
probability flux can reverse direction with continuous
changes in such control parameters as the laser power or
temperature.

Our discussion focuses on the
behavior of a single
particle immersed in a viscous medium that also acts as
a thermodynamic heat bath.
The particle is acted upon by 
a static force field, 
\begin{equation}
  \label{eq:hodge}
  \vec{F}(\vec{r}) = -\nabla U + \nabla \times \vec{A},
\end{equation}
that is factored uniquely
into a conservative irrotational component described
by the scalar potential $U(\vec{r})$ 
and a non-conservative solenoidal component derived
from the vector potential $\vec{A}(\vec{r})$.
In the absence of thermal forcing, the particle is assumed
to come to a stable mechanical equilibrium at a point
$\vec{r}_0$ for which $\vec{F}(\vec{r}_0) = 0$.

\begin{figure}[!t]
  \centering
  \includegraphics[width=\columnwidth]{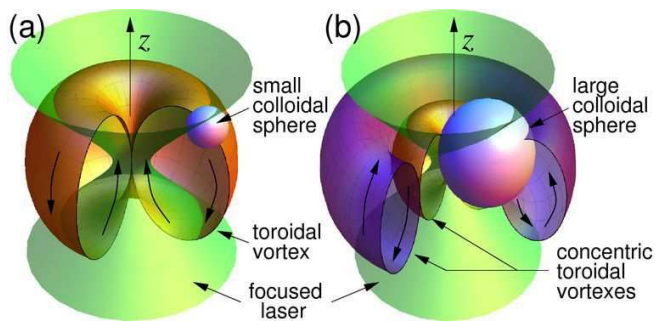}
  \caption{(Color online) 
    Schematic representation of a colloidal sphere undergoing
    toroidal circulation in an optical tweezer.  (a) The Rayleigh
    limit, $a < \lambda$.  (b) The Mie and ray-optics limits,
    $a > \lambda$.}
  \label{fig:schematic}
\end{figure}

Were it not for thermal forces, the particle would remain
motionless at $\vec{r}_0$.
Thermal fluctuations enable the
particle to explore the force field with a probability
density $\rho(\vec{r})$ that does not depend on time
in steady state.
The local probability flux,
\begin{equation}
  \label{eq:flux}
  \vec{j}(\vec{r}) =
  \mu \, \rho(\vec{r}) \, \vec{F}(\vec{r}) - 
  \mu \, k_B T \, \nabla \rho(\vec{r}),
\end{equation}
reflects both the particle's response to the force field
through its mobility, $\mu$, 
and also its diffusivity at temperature $T$.
Because the probability is conserved,
$\nabla \cdot \vec{j} = 0$, and the
flux either vanishes in equilibrium ($\nabla \times \vec{A} = 0$)
or else forms closed loops in steady state
($\nabla \times \vec{A} \neq 0$).
Circulation around these loops is measured
by 
\begin{equation}
  \label{eq:circulation}
  \nabla \times \vec{j} =
  \mu \rho \, \vec{\omega} + \mu \nabla \rho \times \vec{F}, 
\end{equation}
where $\vec{\omega}(\vec{r}) = -\nabla^2\vec{A}$ is the
vorticity in $\vec{F}(\vec{r})$ \cite{morino86}.
Vortexes can form in $\vec{j}(\vec{r})$
even if $\vec{F}(\vec{r})$ does not have
closed loops because the diffusive flux represented
by the second term in Eq.~(\ref{eq:circulation}) 
provides the return flow.

Both the magnitude and the direction 
of $\vec{\omega}(\vec{r})$ can vary with
position.
The rate and direction of the particle's
circulation therefore depend on the domain over which
the particle can diffuse at temperature $T$.
Changing the temperature changes this range and therefore
can reverse the sense of the overall circulation.
The possibility of temperature-dependent
flux reversal distinguishes Brownian vortex circulation
from the more familiar interplay of advection and diffusion
in such systems as the 
electric current flowing through a battery-powered circuit.

As a concrete example of the Brownian vortex phenomenon,
we consider the motions of a colloidal sphere in a
optical tweezer \cite{ashkin86}, a single-beam optical
gradient force trap that consists of a strongly
focused beam of light \cite{ashkin86}.
Recent particle-tracking
measurements \cite{roichman08a} 
overturned the conventional understanding that a trapped
particle comes to thermodynamic equilibrium within an
optical tweezer, revealing instead a circulating steady
state indicated schematically in Fig.~\ref{fig:schematic}(a).
The original explanation
for these observations \cite{roichman08a}
relied on an idealized
model for the optical forces acting on a sphere
in which intensity gradients drew the sphere to the
focal point with a Hookean force and radiation pressure
displaced the particle downstream with a force proportional
to the local intensity.
This model predicts the appearance of a single toroidal
vortex in $\rho(\vec{r})$ directed downstream along the
optical axis and returning upstream at larger displacements.


\begin{figure}[!t]
  \centering
  \includegraphics[width=0.88\columnwidth]{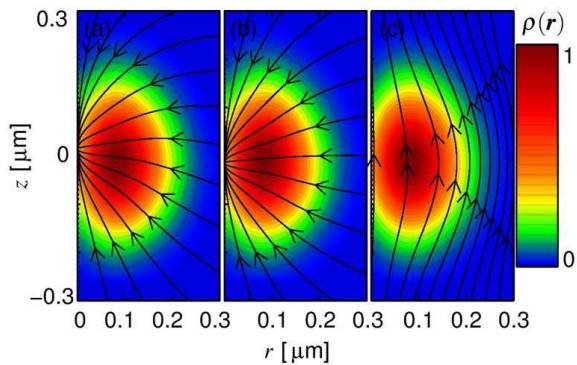}
  \caption{(Color online) (a) Streamlines of 
    the optical force $\vec{F}(\vec{r})$ in the $(r,z)$
    plane exerted on a 0.4~\micron radius
    polystyrene sphere in water by a 532~\unit{nm} optical tweezer
    focused with a numerical aperture 1.4 lens.
    (b) Streamlines of the irrotational
    component of the force, $-\nabla U$.  
    (c) Streamlines of the solenoidal component of the 
    force, $\nabla \times \vec{A}$.
    Shading indicates the relative probability density
    $\rho(\vec{r})$ at
    $T = 21\unit{^\circ C}$ and a laser power of 0.3~\unit{W}.}
  \label{fig:force}
\end{figure}

Light scattering by colloidal particles is more complicated
than suggested by this model, particularly for particles
such as those studied in Ref.~\cite{roichman08a} 
that are larger than the
wavelength of light \cite{bohren83,ashkin92}.
Figure~\ref{fig:force}(a) shows streamlines of the force field computed
with Lorenz-Mie theory \cite{sun08} for
a micrometer-diameter polystyrene sphere in water
trapped within an optical tweezer.
The trap is formed from a laser beam with a vacuum wavelength
of $\lambda = 532~\unit{nm}$ propagating in the $\uvec{z}$ direction and
brought to a focus by an ideal lens with numerical aperture 1.4.
Streamlines are projected into the $(r, z)$ plane
in cylindrical coordinates, $\vec{r} = (r, \phi, z)$.
Figures~\ref{fig:force}(b) and (c) show the irrotational and
solenoidal components of the force field, respectively, and
were obtained through the Helmholtz-Hodge decomposition \cite{morino86}
of $\vec{F}(\vec{r})$.

The particle's thermally driven trajectory $\vec{r}_p(t)$
through this force field was computed
with a Brownian dynamics simulation of the Langevin equation
\begin{equation}
  \label{eq:langevin}
  \dot{\vec{r}}_p = \mu \, \vec{F}(\vec{r}_p) + \mu \, \vec{f}(t),
\end{equation}
where $\mu = (6 \pi \eta a)^{-1}$ is the Stokes mobility for a
sphere of radius $a$ moving through a fluid of viscosity $\eta$,
and where the stochastic force due to thermal
fluctuations satisfies
$\avg{\vec{f}(t)} = 0$ and
$\avg{\vec{f}(t) \cdot \vec{f}(t^\prime)} 
= 2 \mu^{-1} k_B T \, \delta(t - t^\prime)$.
The probability distribution and flux then are computed
with nonparametric density estimators \cite{silverman92}
as
\begin{align}
  \label{eq:densityestimator}
  \rho(\vec{r}) 
  & = 
  \frac{1}{(2\pi\sigma)^{3/2}} \,
  \avg{\exp\left(- \frac{\abs{\vec{r}-\vec{r}_p(t)}^2}{2 \sigma^2} \right)}
  \quad \text{and}\\
  \label{eq:fluxestimator}
  \vec{j}(\vec{r})
  & =
  \frac{1}{(2\pi\sigma)^{3/2}} \,
  \avg{\vec{r}_p(t) \, \exp\left(- \frac{\abs{\vec{r}-\vec{r}_p(t)}^2}{2 \sigma^2} \right)},
\end{align}
where $\sigma$ is chosen to minimize the variance in $\rho(\vec{r})$
without unnecessary blurring.
This approach lends itself to direct comparison between simulation and
experimental measurements of colloidal trajectories.

\begin{figure}[!t]
  \centering
  \includegraphics[width=\columnwidth]{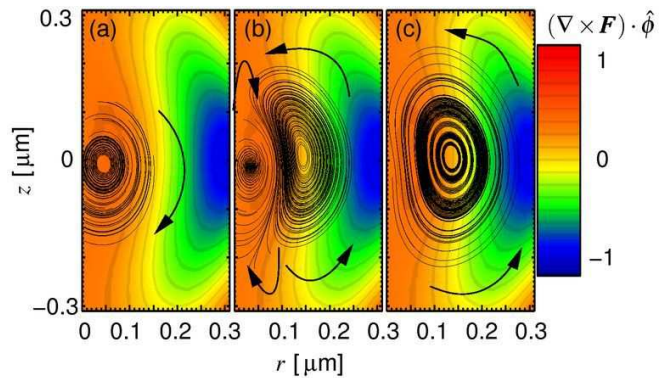}
  \caption{(Color online) 
    Computed streamlines of a 0.4~\micron radius polystyrene sphere
    diffusing in the optical force field of Fig.~\protect~\ref{fig:force}.
    (a) Forward circulation at $P = 0.7~\unit{W}$.
    (b) Concentric counter-rotating vortexes at $P = 0.5~\unit{W}$.
    (c) Complete flux reversal at $P = 0.3~\unit{W}$.}
  \label{fig:streamlines}
\end{figure}
Shading in Fig.~\ref{fig:force} represents the probability
density $\rho(\vec{r})$ at $T = 21\unit{^\circ C}$
for an optical tweezer powered by $P = 0.3~\unit{W}$.
Neither, $\vec{F}(\vec{r})$ nor its solenoidal component
display loops in this range.
Nevertheless, the streamlines of $\vec{j}(\vec{r})$, plotted in
Fig.~\ref{fig:streamlines} show loops in the $(r,z)$ plane
consistent with the appearance of a toroidal vortex centered
on the optical axis.
It should be emphasized that the observed circulation is
apparent only in a very long trajectory or in 
an ensemble average of shorter trajectories; at short times, the particle's
motion resembles a random walk in a harmonic well, its toroidal bias
too subtle to be perceived.

Figure~\ref{fig:streamlines}(a) shows streamlines at a comparatively
large laser power, $P = 0.7~\unit{W}$, 
for which the particle is well localized near the optical axis.
Under these conditions, the particle circulates in a single
toroidal vortex, much as was
predicted in Ref.~\cite{roichman08a} and 
portrayed in Fig.~\ref{fig:schematic}(a).
The local circulation rate,
$[\nabla \times \vec{j}(\vec{r})] \cdot \uvec{\phi}$, 
is uniformly positive.

Reducing the laser power does not change the structure of the
force field, but reduces its overall magnitude.
This is equivalent, therefore, to increasing the effective
temperature.
Doing so increases the range over which the particle can wander
and enables it to populate a second, concentric counter-rotating
vortex, as plotted in Fig.~\ref{fig:streamlines}(b) for
$P = 0.5~\unit{W}$, and indicated
schematically in Fig.~\ref{fig:schematic}(b).

At still lower laser power (or higher temperature), the outer vortex
supplants the inner vortex, and the probability current circulates
once again in a single toroidal roll, but with its direction
reversed.
This flux reversal
is demonstrated in Fig.~\ref{fig:streamlines}(c) for $P = 0.3~\unit{W}$.

The origin of the flux reversal in Brownian vortex circulation
can be seen in the vorticity
of the force field
in Fig.~\ref{fig:streamlines}.
Although the solenoidal component of the optical force,
$\nabla \times \vec{A}$,
is directed uniformly upward, $\nabla \times \vec{F}$
changes direction with distance from the optical axis.
So long as the particle's
probability density is concentrated in regions where
$\vec{\omega}(\vec{r})\cdot \uvec{\phi}$ is positive, as is the
case in Fig.~\ref{fig:streamlines}(a),
the overall circulation of the probability flux also is positive.
When the particle wanders into regions of negative
vorticity, it circulates in the retrograde direction, as shown
in Fig.~\ref{fig:streamlines}(b).
In both cases, the non-conservative part of the optical force field
redistributes $\rho(\vec{r})$ downstream of the beam's focal point
and diffusion provides the return current.
The single-roll structure reasserts itself in the flux-reversed state
when gradients in $\rho(\vec{r})$ become large enough for diffusion
to outstrip advection along the optical axis.

We also observed flux reversal in Brownian vortex circulation
through experimental observations of colloidal 
spheres trapped in optical tweezers.
Our system consists of 1.5~\unit{nm} diameter 
colloidal silica spheres (Bangs Labs, Lot SS04N/5252)
dispersed in a 50~\micron thick layer of water that is hermetically
sealed between a clean glass slide and a No.~1.5 cover slip.
The sample is mounted on the stage of an inverted light microscope
(Nikon TE 2000U) where it is observed with a $100\times$ numerical
aperture 1.4 oil immersion objective lens (Nikon Plan Apo).
The same objective lens is used to focus four holographic
optical tweezers \cite{dufresne98,grier03,polin05,roichman06a}
arranged at the corner of a square with 30~\micron
sides near the midplane of the sample.
These traps are powered by a single laser (Coherent Verdi 5W)
operating at a vacuum wavelength of $\lambda = 532~\unit{nm}$
that is imprinted with a computer-generated hologram \cite{polin05}
by a liquid-crystal spatial light modulator
(Hamamatsu X7665-16) before being projected into the sample.
The trap array is designed so that two of the traps have
nearly the same intensity, the third is slightly brighter, and
the fourth brighter still.
This enables us to seek out intensity-dependent differences
in simultaneously acquired data sets, and thus to avoid possible
systematic effects due to vibrations or other instrumental
fluctuations.

\begin{figure}
  \centering
  \includegraphics[width=\columnwidth]{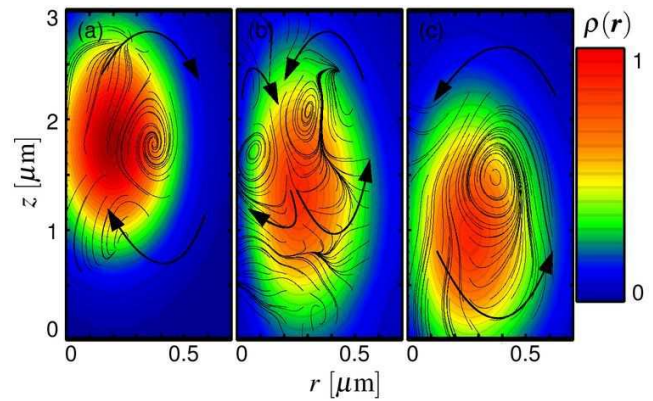}
  \caption{(Color online) Streamlines of the measured circulation of 
    colloidal silica spheres trapped in optical tweezers.
    (a) $k_\perp = 6.56~\unit{pN/\micron}$: Single roll circulating in the
    positive sense.
    (b) $k_\perp = 2.43~\unit{pN/\micron}$: Concentric counter-rotating
    rolls.
    (c) $k_\perp = 2.27~\unit{pN/\micron}$: Flux reversal.  Single
    retrograde roll.
    Background images show the particles' measured probability
    densities.
  }
  \label{fig:expt}
\end{figure}

The trapped spheres' three-dimensional motions are measured
with nanometer resolution
through quantitative analysis \cite{lee07a} of images
obtained with holographic video microscopy
\cite{sheng06,lee07}.
Holographic images are obtained by illuminating the sample
with the collimated beam from
a HeNe laser (Uniphase 10~\unit{mW}) operating at 632.8~\unit{nm}.
Light scattered by the particles interferes with
the unscattered portion of the beam in the microscope's
focal plane to create an in-line hologram that is magnified
and recorded by a video camera (NEC TI-324AII) at 30 frames per
second.
Radiation pressure due to the 
$\unit{nW/\micron^2}$ intensity of the imaging
beam is negligible compared with thermal forces and
forces due to the optical trap and so does not affect
the particles' trajectories.

Both fields of each interlaced holographic video frame 
were analyzed with Lorenz-Mie
light scattering theory \cite{bohren83,lee07a}
to measure each sphere's three-dimensional position
$\vec{r}(t)$, with 3~\unit{nm} in-plane resolution and
10~\unit{nm} axial resolution \cite{lee07a}
A total of thirty two-minute-long trajectories were acquired 
at constant laser power for the four particles.
At the end of each acquisition period, the trapped particles
were moved automatically out of the field of view to
acquire background holograms and to confirm the system's
stability.
Each particle's trajectories were analyzed with
Eqs.~(\ref{eq:densityestimator}) and (\ref{eq:fluxestimator})
to visualize the mean circulation
and the results combined into maps of the mean circulation
for each trap.
In all, more than 100,000 holograms were analyzed for each trap.

Figure~\ref{fig:expt} shows streamlines of the trajectories
for three of the four particles, the fourth serving as a control
for Fig.~\ref{fig:expt}(c).
These results confirm not only the presence of toroidal circulation
in the particles' motions, but also the appearance of flux
reversal as a function of trap strength.
Each trap-particle combination 
is characterized by its apparent \cite{roichman08a}
in-plane stiffness, $k_\perp$, which is obtained from
statistical analysis of the particle's
measured in-plane fluctuations \cite{polin05}, stiffer
traps generally corresponding to higher laser power
and lower effective temperature.
The traps' apparent axial stiffness is a factor of 5 smaller
than their lateral stiffness because axial intensity gradients
are correspondingly weaker \cite{roichman08a,vermeulen06}.

The stiffest trap, shown in Fig.~\ref{fig:expt}(a), concentrates
its particle's probability density $\rho(\vec{r})$ closest to the
optical axis and displays a single roll circulating in the positive
sense.
The weaker trap in Fig.~\ref{fig:expt}(b) allows the
trapped particle to wander further afield, where it enters
into concentric counter-rotating rolls, similar to the
simulated results in Fig.~\ref{fig:streamlines}(b).
The weakest traps, one of which is represented in Fig.~\ref{fig:expt}(c),
both display a single retrograde roll in $\vec{j}(\vec{r})$,
and thus demonstrate complete flux reversal.

The probability distribution is centered lower in the weaker
traps because of gravity acting on the silica spheres, whose
$1.9~\unit{g/cm^3}$ density exceeds that of the
surrounding water.
This additional conservative force does not directly
contribute to the particles' circulation but does affect 
what region of the optical force field the particle occupies
for a given laser power.
Undoubtedly, this influenced the trend in Fig.~\ref{fig:expt}, but
does not change our interpretation of the phenomenon as
flux reversal in a Brownian vortex.

Instrumental fluctuations cannot account for our
observations because all four measurements were performed
simultaneously in a static array of optical traps derived
from the same laser beam.
The particles are sufficiently separated from each other
and from the walls of their container that hydrodynamic
coupling also is unlikely to have influenced their motion
\cite{hydrodynamics}.
Rather, Brownian vortex circulation, including power-dependent
flux reversal, appears to be an inherent aspect of the
statistics of colloidal spheres in optical tweezers.

In this study, we have introduced the Brownian vortex as
a distinct class of noise-driven machines.
Unlike stochastic heat engines driven by time-dependent forces,
Brownian vortexes are driven out of equilibrium
by the solenoidal component of a static force field.
Because one-dimensional force fields have no solenoidal component,
the Brownian vortex has no one-dimensional manifestation.
Not any static force field, furthermore,
can support a Brownian vortex.
For example, a force field lacking a stable equilibrium point
cannot establish the requisite probability-conserving steady-state.
Still other force fields
establish circulating steady states without thermal noise.
An example of this is provided by the ring-like optical
trap known as an optical vortex \cite{vortex} 
that exerts torques on trapped objects \cite{he95a}
through its helical wavefront structure \cite{phaseforce}.
These are deterministic machines rather than 
stochastic heat engines, and so are not Brownian
vortexes.

Although the illustrative simulations and experiments presented
here focus on colloidal circulation in optical tweezers,
the Brownian vortex is a more general phenomenon.
Seeking its signature in such contexts as biological
networks and financial systems, as well as in new mechanical
models, should provide opportunities for future research.
Further work also is required to elucidate Brownian
vortexes' thermodynamic properties, particularly 
the considerations that determine their
thermodynamic efficiency.

We acknowledge fruitful conversations with Yohai Roichman.
This work was supported by the MRSEC program of the NSF
under grant number DMR-0820341.
B.S. acknowledges support from the Kessler Family Foundation.


\end{document}